\documentclass{webofc}
\usepackage[varg]{txfonts}   % Web of Conferences font
\begin{document}
\title{The reaction $\pi N \to \omega N$ in a dynamical coupled-channel approach}
\author{\firstname{Yu-Fei} \lastname{Wang}\inst{1}\fnsep\thanks{\email{yuf.wang@fz-juelich.de}}}

\institute{Institute for Advanced Simulation and J\"ulich Center for Hadron Physics, Forschungszentrum J\"ulich, 
52425 J\"ulich, Germany}

\abstract{This talk is on a refined investigation on light flavor meson-baryon scatterings, using a dynamical coupled-channel approach, i.e. the Jülich-Bonn model. The previous channel space of $\pi N$, $\pi \Delta$, $\sigma N$, $\rho N$, $\eta N$, $K \Lambda$ and $K \Sigma$ is extended by adding the $\omega N$ final state. The spectra of $N^*$ and $\Delta$ resonances are extracted, based on the result of a global fit to a worldwide collection of data, in the energy region from the $\pi N$ threshold to center-of-mass energy $z=2.3$ GeV (approximately $300$ parameters against $9000$ data points). A negative value of the $\omega N$ elastic spin-averaged scattering length has been extracted. }
\maketitle
The spectra of $N^*$ and $\Delta$ particles are usually extracted from the scattering of light mesons and baryons, which has been a topic of interest for many decades and continues to be so. In the high energy region, the interaction can be treated perturbatively, whereas, when the energy is very low, effective field theories~\cite{Meissner:2022cbi} can be considered. However, though there are abundant experimental observations, the dynamics in the middle energy region is very challenging, and conventional Breit-Wigner descriptions of the cross sections may fail due to the involved dynamics. Partial-wave analyses (PWA) is a common way to extract resonances. Among different methods of PWA, dynamical coupled-channel (DCC) models provide a sophisticated tool on this problem, since the partial-wave amplitudes are solved via dynamical scattering equations, respecting the analytical properties and defining the resonances as the complex poles on the second Riemann sheet. The J{\"u}lich-Bonn model is a DCC model with decades of development, which now covers the energy region from $\pi N$ threshold to $2.3$ GeV and fits to a worldwide collection of data. Its central part is the hadronic model~\cite{schuetz1994,schuetz1995,schuetz1998,Krehl2000,Gasparyan2003,Doering2009,Doering2011,Roenchen2013,Wang2022} that contains the interaction channels induced by $\pi N$. This model has also been applied to the photoproduction~\cite{Roenchen2014,Roenchen2015,Roenchen2018,Roenchen2022} and electroproduction~\cite{Mai2021,Mai2022,Mai2023} of the mesons. The same method can also be used to study the $P_c$ states~\cite{Shen2018,ZLWang2022}. 

This talk is based on a recent work~\cite{Wang2022}: the inclusion of the $\omega N$ channel in the hadronic part of this model. Due to vector meson dominance~\cite{VMD}, the $\omega N$ interaction is crucial for determining the properties of the nuclear matter via dilepton emission. It has also been found that the $\omega$ meson plays a very important role in the equation of state of the neutron stars~\cite{Shen1998}. Basically, the real part of the $\omega N$ elastic scattering length gives hints whether or not in-medium bound states of $\omega$ tend to be formed. Experiments cannot directly measure the $\omega N$ elastic scatterings, so the scattering length should be reliably extracted from DCC models that are well-constrained by a large amount of data points. Apart from providing the detailed $N^*$ and $\Delta$ spectrum, the J{\"u}lich-Bonn model is a very good candidate to investigate the $\omega N$ scattering length. 

The dynamics in this model is described by the Lippmann-Schwinger-like equation: 
\begin{equation}\label{scequ}
  T_{\mu\nu}(p'',p',z)=V_{\mu\nu}(p'',p',z)
  +\sum_{\kappa}\int_0^\infty p^2 dp V_{\mu\kappa}(p'',p,z)G_{\kappa}(p,z)T_{\kappa\nu}(p,p',z)\ ,
\end{equation}
where $p'$ ($p''$) stands for the initial (final) state momentum, $z$ is the centre-of-mass energy, $V$ is the potential with various hadron exchange diagrams, $T$ is the amplitude and $G$ is the propagator. The Greek letters are the channel labels for $\pi N$, $\pi\pi N$ (simulated by effective channels $\sigma N$, $\rho N$, and $\pi\Delta$), $\eta N$, $K\Lambda$, $K\Sigma$, and $\omega N$, with specific isospin ($I$), angular momentum ($J$, up to $9/2$), spin ($S$) and orbital angular momentum ($L$). The propagator is $G_{\kappa}=(z-E_\kappa-\omega_\kappa-\Sigma_\kappa)^{-1}$, with $E_\kappa$ and $\omega_\kappa$ being the energies of the baryon and meson in channel $\kappa$, respectively; $\Sigma_\kappa$ is the self-energy of the unstable particle ($\sigma$, $\rho$, and $\Delta$) when $\kappa$ is an effective channel. For technical details, see Ref.~\cite{Wang2022} and its supplemental material. 

\begin{figure}[t]
\centering
\vspace*{0.1cm}
\includegraphics[width=6.5cm,clip]{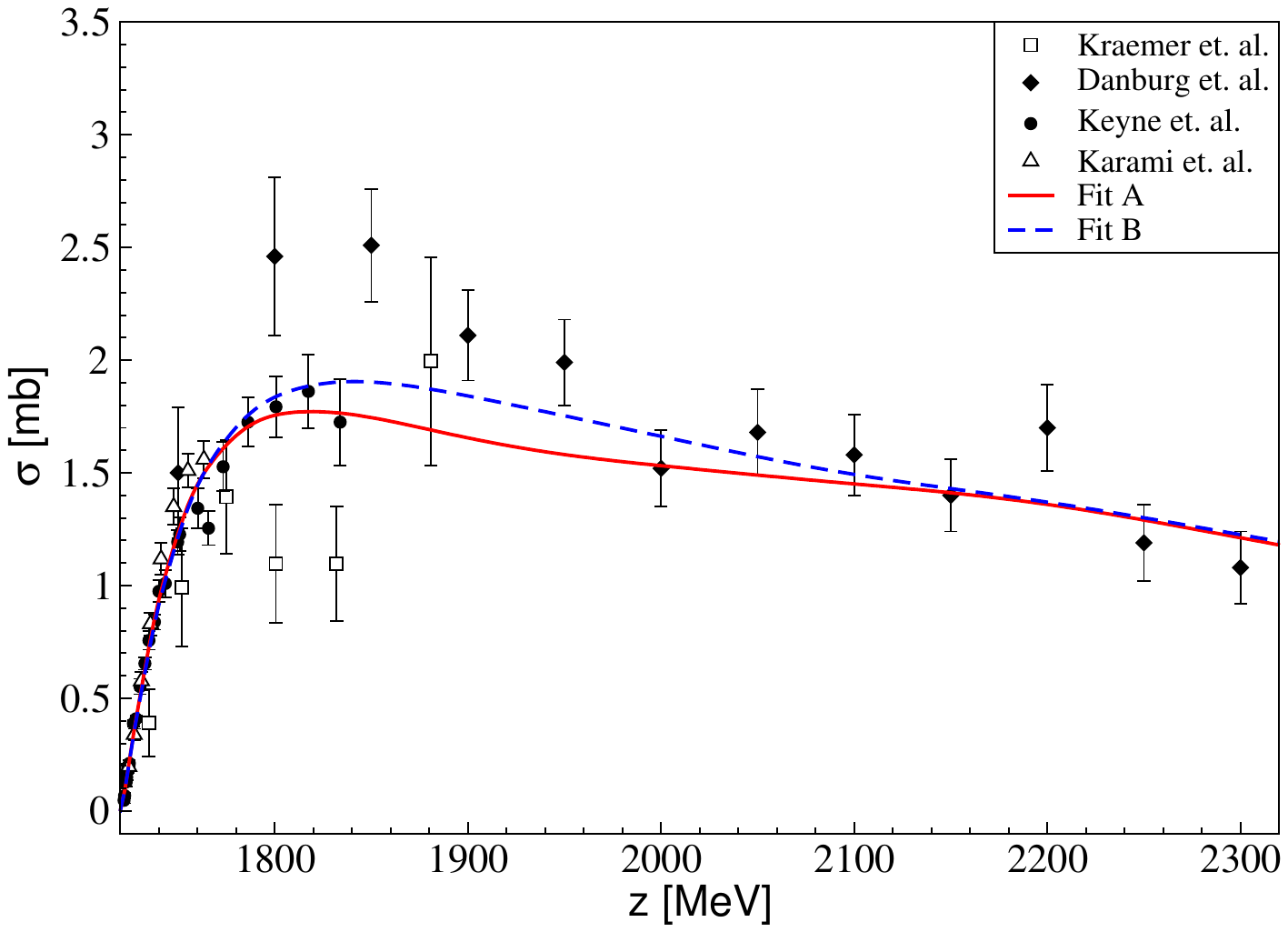}
\caption{The total cross section of the $\pi N\to\omega N$ reaction, with $z$ being the centre-of-mass energy. The data points are from Refs.~\cite{data1,data2,data3,data4}. }
\label{fig:TXS}      
\end{figure}

The free parameters are all in the potential $V$: cut-off parameters of the regulators, bare masses and couplings of the $s$-channel intermediate states. In this study $304$ fit parameters are adjusted using $9000$ data points, among which $174$ belong to the $\omega N$ reaction. The fit is performed on the supercomputer JURECA~\cite{JU} at the Forschungszentrum J{\"u}lich. However, such a complicated fit cannot be done without adding special weights to the $\omega N$ data. Therefore, statistical errors are hard to be addressed. Two different fit scenarios have been implemented to estimate the uncertainties: fit A starts from intermediate parameter values of Ref.~\cite{Roenchen2018}, while fit B starts from another set of parameter values which contains an extra narrow pole in the $\pi N$ $P_{11}$ wave. They both fit the data well globally, e.g. see Fig.~\ref{fig:TXS}. 

\begin{figure}[t]
\centering
\vspace*{0.1cm}
\includegraphics[width=6.5cm,clip]{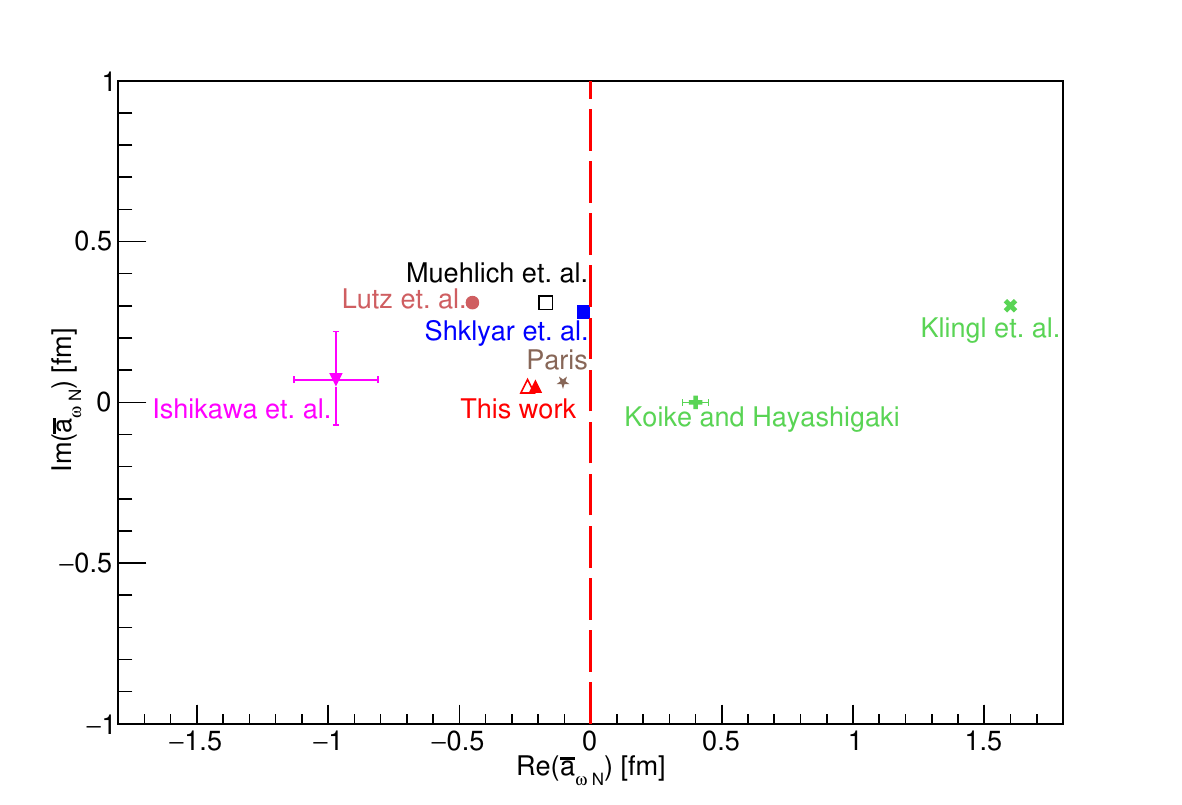}
\caption{A summary of the spin-averaged $\omega N$ scattering length extractions. The result from fit~A~(B) is denoted by the empty (filled) red triangle. The others are results from Refs.~\cite{Koike:1996ga,Klingl:1998zj,Lutz:2001mi,Shklyar:2004ba,Muehlich:2006nn,Paris:2008ig,Ishikawa:2019rvz}. }
\label{fig:oNasc}      
\end{figure}

In this study, $17$ $N^*$ poles and $15$ $\Delta$ poles have been found; for details see Ref.~\cite{Wang2022}. The residues of $N(1535)\,\frac{1}{2}^-$, $N(1710)\,\frac{1}{2}^+$ and $N(1680)\,\frac{5}{2}^+$ in the $\omega N$ channel are relatively larger, which indicates the stronger coupling strengths of these lower states to $\omega N$. In fact the bare couplings of the $N(1535)\,\frac{1}{2}^-$ and $N(1710)\,\frac{1}{2}^+$ bare states to the $\omega N$ channel are very large. We have attempted to perform a ``fit C'' with constraints on the bare couplings but have failed. The spin-averaged scattering length of $\omega N$ is extracted: $(-0.24+0.05i)$ fm in fit A and $(-0.21+0.05i)$ fm in fit B. Both fits result in a negative real part, which does not favour the in-medium bound states of $\omega$. Fig.~\ref{fig:oNasc} is a comparison of the scattering lengths obtained in this work and the others; most of the results support a negative real part. In addition, in Ref.~\cite{Strakovsky} the absolute value of the scattering length is evaluated as $|\bar{a}_{\omega N}|=0.82\pm 0.03$ fm . 

To summarize, this talk reports a recent work on the study of $\omega N$ interaction using the J{\"u}lich-Bonn model, a powerful tool for the partial-wave analyses and extraction of the hadron spectra. More than $9000$ data points in $\pi N$ induced reactions are fitted with two fit solutions estimating the uncertainties. It has been found that $\omega N$ couples strongly to three lower states: $N(1535)\,\frac{1}{2}^-$, $N(1710)\,\frac{1}{2}^+$ and $N(1680)\,\frac{5}{2}^+$. The real part of the $\omega N$ scattering length is negative in this study and the in-medium bound states of $\omega$ are disfavoured. In the future this model can be extended to the $\omega$ photoproduction. Moreover, this model can also be used to study the structures of the resonances~\cite{Wang2023}. 

I would like to thank the collaborators: Deborah R{\"o}chen, Ulf-G.~Mei{\ss}ner, Yu Lu, Chao-Wei Shen, and Jia-Jun Wu. The computing time granted by the JARA Vergabegremium and provided on the JARA Partition part of the
supercomputer JURECA~\cite{JU} at Forschungszentrum J{\"u}lich is acknowledged. This work is supported by the NSFC and the Deutsche Forschungsgemeinschaft (DFG, German Research Foundation) through the funds provided to the Sino-German Collaborative Research
Center TRR110 “Symmetries and the Emergence of Structure in QCD” (NSFC Grant No. 12070131001, DFG Project-ID 196253076-TRR 110) and by the National Natural Science Foundation of China under Grants No. 12175239 and the National Key R\&D Program of China under Contract No. 2020YFA0406400. Further support by the CAS through a President’s International Fellowship Initiative (PIFI)
(Grant No. 2018DM0034) and by the VolkswagenStiftung (Grant No. 93562) is acknowledged.

\end{document}